# AI-based, automated chamber volumetry from gated, non-contrast CT

Athira J Jacob MSc [a], Ola Abdelkarim MD[b], Salma Zook MD [b], Kristian Hay Kragholm, MD, PhD[b], Prantik Gupta BTech[a], Myra Cocker PhD [c], Juan Ramirez Giraldo PhD [c], Jim O Doherty PhD [c], Max Schoebinger PhD [d], Chris Schwemmer MSc[d], Mehmet A Gulsun[a] MSc, Saikiran Rapaka PhD[a], Puneet Sharma PhD [a], Su-Min Chang MD[b*]

[a] Digital Technology and Innovation, Siemens Healthineers, Princeton, NJ

[b] Houston Methodist Hospital, Houston, TX

[c] CT R&D Collaborations, Siemens Healthineers, Malvern, PA

[d] Siemens Healthineers, Forchheim, Germany

**ABSTRACT**

**Background** Accurate chamber volumetry from gated, non-contrast cardiac CT (NCCT) scans can be useful for potential screening of heart failure.

**Objectives:** To validate a new, fully automated, AI-based method for cardiac volume and myocardial mass quantification from NCCT scans compared to contrasted CT Angiography (CCTA).

**Methods:** Of a retrospectively collected cohort of 1051 consecutive patients, 420 patients had both NCCT and CCTA scans at mid-diastolic phase, excluding patients with cardiac devices. Ground truth values were obtained from the CCTA scans.

**Results:** The NCCT volume computation shows good agreement with ground truth values. Volume differences [95% CI ] and correlation coefficients were: -9.6 [-45; 26] mL, r = 0.98 for LV Total, -5.4 [-24; 13] mL, r = 0.95 for LA, -8.7 [-45; 28] mL, r = 0.94 for RV, -5.2 [-27; 17] mL, r = 0.92 for RA, -3.2 [-42; 36] mL, r = 0.91 for LV blood pool, and -6.7 [-39; 26] g, r = 0.94 for LV wall mass, respectively. Mean relative volume errors of less than 7% were obtained for all chambers.

**Conclusions:** Fully automated assessment of chamber volumes from NCCT scans is feasible and correlates well with volumes obtained from contrast study.

## TOC SUMMARY

Cardiac chamber volume is known to be an excellent predictor of future heart failure events. According to recent guidelines, gated, non-contrast cardiac CT (NCCT) is recommended for CAD screening in asymptomatic individuals at low to intermediate risk. We validated a novel, fully automated, AI based method for cardiac chamber volume and myocardial mass quantification from gated NCCT scans used for coronary calcium scoring. NCCT based volume quantification shows excellent agreement with the volumes obtained from CCTA scans in the same patient. Thus, volume assessment from NCCT scans acquired for calcium scoring appears feasible for potential screening.

## ABBREVIATIONS

CAD: coronary artery disease

CT: computed tomography

CCTA: coronary computed tomography angiography

NCCT: non-contrast gated computed tomography

CACS: coronary artery calcium score

AI: artificial intelligence

LV/RV: left/right ventricle

LA/RA: left/right atrium

## 1. Introduction

Coronary artery calcium score (CACS) measurement with non-contrast enhanced, gated cardiac CT (NCCT) scan is used for the detection and prediction of incident CAD in asymptomatic individuals at risk[1]. On the other hand, there is no routine clinical screening for patients who are asymptomatic but at risk of heart failure despite current evidence in favor of early initiation of therapy.[2,3] Large epidemiologic studies using gated NCCT scans for CAD screening have shown that non-coronary data such as LV size and LA size contained in NCCT add further prognostic information.[4,5] There is evidence that cardiac chamber volumes are strong predictors of HF.[6] However, the measurements were based on manual tracing of LA and LV area on a single axial scan which is operator dependent, time consuming and raise concerns for accuracy and reproducibility. While gated NCCT scans are widely used for CAD screening, currently available cardiac CT segmentation algorithms require contrast enhancement for accurate volume assessment. In this work, we build on a previously developed AI based cardiac segmentation algorithm to measure chamber volumes and LV mass from the gated NCCT scans used for calcium scoring.[7] In addition, we sought to evaluate the measured AI volumes and mass in an independent, consecutive clinical cohort, against values derived from gated CCTA as reference.

## 2. Methods

*2.1 AI Algorithm*

The AI algorithm consisted of a 3D Image to Image (I2I) segmentation network, combined with a conditional Variational Autoencoder (cVAE)[7,8] was trained using a cohort of multi-center, paired CCTA and NCCT scans from 2594 patients, to segment left ventricle blood pool (LV BP), myocardium, right ventricle (RV), left atrium (LA), right atrium, (RA) and background. The ground truth segmentation for each NCCT scan were generated from the corresponding CCTA scans[9,10], chosen to be closest in cardiac phase to minimize volume differences. The contours were aligned to the NCCT scan using non-rigid registration algorithms[11]. More details on the algorithm are given in the Supplemental Materials (S.1-3). Algorithm development was performed independently to the clinical validation reported in this work, with no overlap in terms of data or institutions.

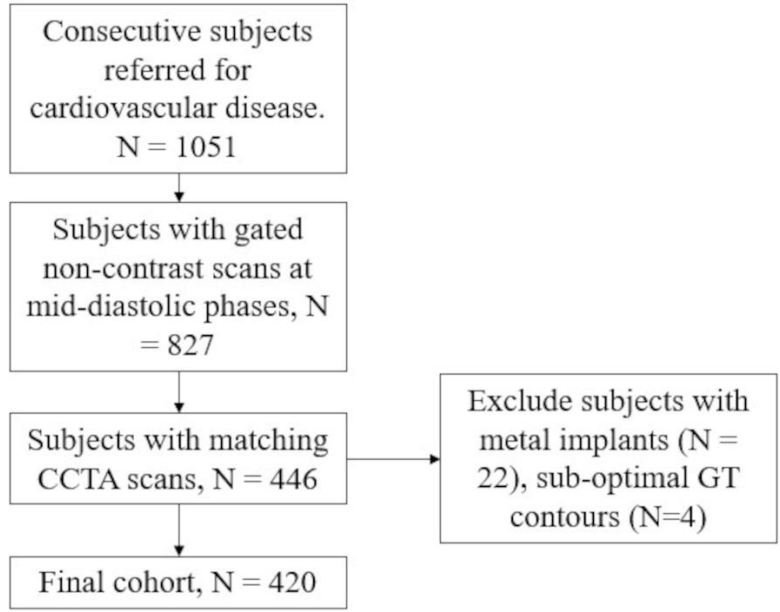

**Figure 1 Cohort selection/exclusion criteria. Flow diagram of cohort selection criteria and reasons for exclusion**

*2.2 Inclusion-exclusion criteria:* Of a cohort of 1051 consecutive patients who underwent CCTA scans for various cardiovascular indications, 827 patients had gated, mid-diastolic NCCT scans performed prior to the contrast study, of which 446 had CCTA scans at matching cardiac phases. After manual review by an expert, 420 patients were included in the final cohort for this study, after excluding patients with metal implants (n=22) and sub-optimal CCTA contours (n = 4) (Figure 1).

*2.3 CT Acquisition:* In this validation study, we evaluated the algorithm on a retrospectively collected cohort of 1051 consecutive patients who underwent contrast enhanced cardiac CT scans for various cardiovascular indications. Acquisitions were performed using a dual-source MDCT system (SOMATOM Force, Siemens Healthineers, Erlangen, Germany). Scan modes included both retrospective, prospective, and dual-source high-pitch ('turbo flash') modes. For contrast-enhanced examinations, typical contrast injection protocol included: visipaque 350, 65 mL injected at 5 mL/s for patients with BMI ≤30, 80 mL injected at 6 mL/s for patients with BMI >30, and 90 mL injected at 7 mL/s in patients with BMI >40. In all cases, contrast was followed with 70-100mL or more saline chaser with injection rate identical to that of the iodinated contrast. CT acquisition parameters including tube potential were automatically selected by the scanner using automated kV selection technology (CAREkV), which automatically picked kV ranging from 70 to 150kV according to patient body habitus. Images were reconstructed with slice thickness of 0.6 mm and 0.4 mm overlap, using medium smooth kernels (Bv36), and advanced modeled iterative reconstruction (ADMIRE) with a strength of 2. The resolution within a slice is isotropic at 0.53 mm. All non-contrast CT acquisitions used a tube potential of either 120, 100Sn or 150Sn kV triggered at mid-diastolic phase, which reflected the institution standard of care for NCCT examinations. Image reconstruction parameters were fixed to quantitative medium smooth kernels (Qr36 and Sa36), at 3.0mm thickness.

*2.4 Study Design*: For the final cohort, the NCCT scan at the mid-diastolic phase was processed by the AI algorithm to segment chambers and myocardium. Volumes were calculated from the segmentations. Myocardial mass was calculated using myocardial specific gravity, 1.05. GT values were obtained from the corresponding contrasted scan acquired within 10% of cardiac phase relative to the NCCT scan. For creating the GT in a fast and reproducible way, we used a previously validated algorithm for chamber segmentation in contrasted scans[10], followed by independent, blinded review of the GT contours by an expert cardiologist to ensure accuracy.

*2.5 Statistical analysis:* Continuous data was reported using mean and standard deviation (SD) and categorical data using counts and percentages. Pearson correlation coefficient was used to measure linear correlation and Bland Altman analyses to assess distribution of quantitative differences. A p-value of <0.05 was considered significant. All statistical analyses were done using MedCalc version 20.015.

3. Results

**Table 1. Patient cohort demographics. BSA = body surface area; CAD = coronary artery disease. Clinical variables were available for 417 of the 420 patients.**

| **General** | |
|---|---|
| Age, years | 63 ± 13 |
| Gender, female | 185 (44%) |
| Height (m) | 1.7 ± 0.1 |
| Weight (kg) | 85.7 ± 21.3 |
| BSA (m²) | 1.97 ± 0.26 |
| **Study Indications** | |
| CAD Assessment | 314 (75%) |
| Other | 103 (25%) |

**Table 2: Mean ( ± SD) chamber volumes (or mass), and indexed volumes (or mass) from CCTA in the cohort. Volumes are in mL, mass in g. SD = standard deviation, BP = blood pool**

| Structure | Volumes (mL or g) | Volume Indices (mL/m² or g/m²) |
|---|---|---|
| LV Total | 264.1 ± 83.4 | 132.0 ± 34.7 |
| LV BP | 133.7 ± 47.3 | 67.0 ± 20.8 |
| LV Mass | 136.9 ± 43.9 | 68.2 ± 18.1 |
| RV | 170.0 ± 52.3 | 85.1 ± 20.2 |
| LA | 87.2 ± 29.4 | 43.7 ± 15.0 |
| RA | 80.2 ± 29.1 | 40.3 ± 13.6 |

Demographic data and cohort chamber volumes are summarized in Tables 1 and 2. Example segmentations are shown in Figure 2. There was excellent agreement between the CCTA and NCCT volumes, with correlations above 0.9, and mean absolute and relative errors of less than 10 mL and 7% respectively, for every chamber. LV Total and LA volumes had the best correlations of 0.98 and 0.95 respectively, with absolute biases and limits of agreements (LOAs) of -9.6 [-45, 26] mL, and -5.4 [-24, 13] mL. RV and RA volumes had absolute biases of -8.7 [-45,28] mL and -5.2 [-27,17] mL, with correlations of 0.94 and 0.92 respectively. The reference ranges provided in Fuchs et al[12], were used to define patients having at least two chambers with volumes beyond 2 standard deviations as abnormal. High correlations are observed in the abnormal sub-group, ranging from 0.89 for LV BP to 0.97 for LV Total. The correlations, means and LOAs are given in Table 3 and Figure 3. The NCCT volumes were able to distinguish normal and abnormal chambers with high AUCs ( LV BP: 0.96, LV Mass: 0.96, RV: 0.97, LA: 0.97, RA: 0.99).

Reproducibility was measured by rerunning the algorithm on the entire cohort. Correlations of 1.0 were obtained for all chamber volumes, and myocardial mass, demonstrating perfect reproducibility of the algorithm. Automated segmentation for all structures took 20 s per dataset on a machine with Nvidia Titan X 12 GB graphics card.

Table 3: Quantitative assessment: Bland Altman biases, limits of agreements (LOA), correlations, p values. BP = blood pool

| Structure | Bland-Altman Bias (Absolute) [LOA] (mL or g) | Bland-Altman Bias (Relative) [LOA] (%) | Pearson's Correlation |
|---|---|---|---|
| LV Total | -9.6 [-45, 26], $p < 0.0001$ | -3.6 [-17, 10], $p < 0.0001$ | 0.98, $p < 0.0001$ |
| LV BP | -3.2 [-42, 36], $p = 0.0010$ | -2.5 [-33, 28], $p = 0.0012$ | 0.91, $p < 0.0001$ |
| LV Mass (g) | -6.7 [-39, 26], $p < 0.0001$ | -3.7 [-24, 16], $p < 0.0001$ | 0.94, $p < 0.0001$ |
| RV | -8.7 [-45, 28], $p < 0.0001$ | -5.9 [-29, 17], $p < 0.0001$ | 0.94, $p < 0.0001$ |
| LA | -5.4 [-24, 13], $p < 0.0001$ | -6.8 [-28, 15], $p < 0.0001$ | 0.95, $p < 0.0001$ |
| RA | -5.2 [-27, 17], $p < 0.0001$ | -6.3 [-34, 21], $p < 0.0001$ | 0.92, $p < 0.0001$ |
| Sub-group: Abnormal (N = 121) | | | |
| LV Total | -15.4 [-63, 32], $p < 0.0001$ | -4.2 [-20, 11], $p < 0.0001$ | 0.97, $p < 0.0001$ |
| LV BP | -2.6 [-55, 50], $p = 0.29$ | -0.2 [-37, 37], $p = 0.91$ | 0.89, $p < 0.0001$ |
| LV Mass (g) | -13.5 [-53, 26], $p < 0.0001$ | -6.9 [-28, 14], $p < 0.0001$ | 0.90, $p < 0.0001$ |
| RV | -6.4 [-54, 41], $p < 0.004$ | -2.7 [-27, 21], $p < 0.018$ | 0.92, $p < 0.0001$ |
| LA | -6.5 [-32, 19], $p < 0.0001$ | -5.9 [-30, 18], $p < 0.0001$ | 0.93, $p < 0.0001$ |
| RA | -8.0 [-36, 20], $p < 0.0001$ | -7.3 [-36, 21], $p < 0.0001$ | 0.92, $p < 0.0001$ |

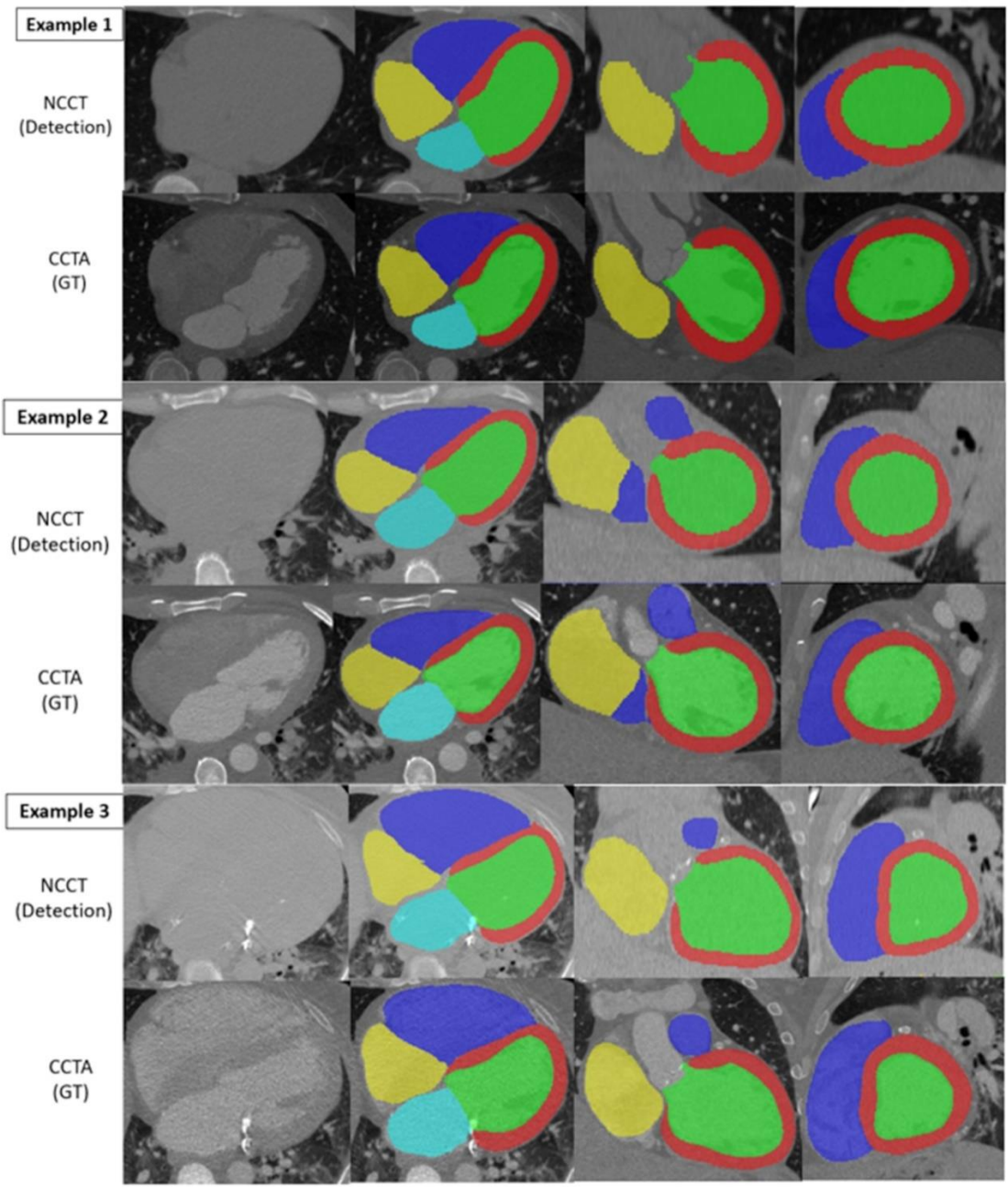

Figure 2 Segmentations for 3 patients with varying total heart volumes of 438 mL, 814 mL, 1551 mL) respectively. CCTA = coronary computed tomography angiography. GT = ground truth. NCCT = Non contrast computed tomography

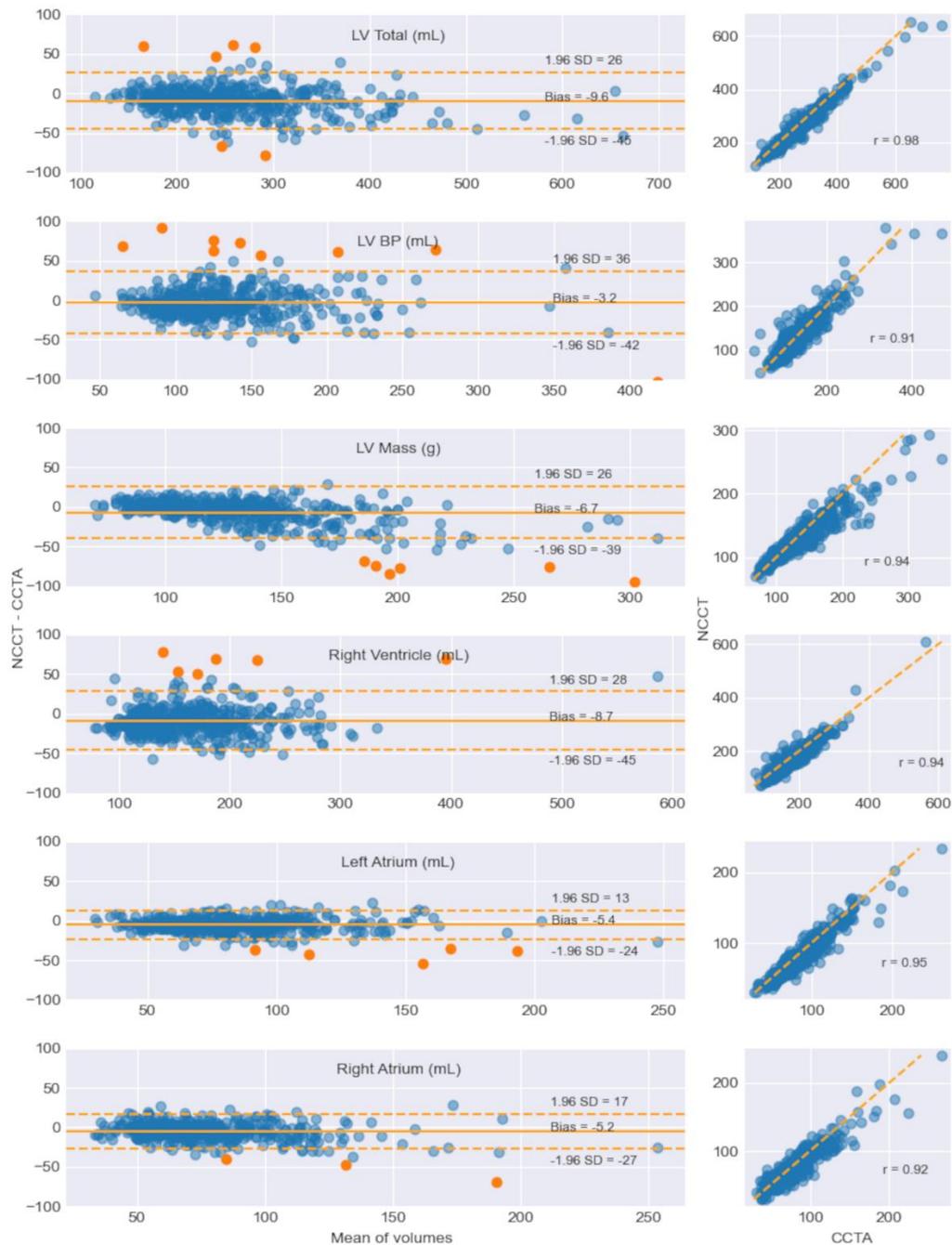

**Figure 3 Bland Altman and correlation plots for detected NCCT volumes versus GT . Solid orange line in Bland Altman plots depict mean absolute difference, and dashed lines represent 95% confidence interval Dashed orange lines in scatter plots represent the identity line. r = Pearson's correlation coefficient. SD = standard deviation. Other abbreviations as in Figure 1.**

4. **Discussion**

Coronary artery calcium scoring (CACS) on native, non-contrast scans is widely used as a screening test for CAD. Cardiac chamber volume analysis from the same NCCT scans can provide an excellent opportunity to screen for early heart failure (stage B) at no additional cost or burden. We developed and validated a fully-automated, AI-based model for multi-chamber cardiac volumetry from NCCT scans, achieving high accuracy. The study used a large, consecutive clinical cohort with a wide range of volumes, with up to twice the SD when compared with reference values from a normal population.[12] The study demonstrated excellent positive correlation, exceeding 0.9, for all chamber volumes particularly for the left atrial and LV total volumes.

There are a few studies on automatic segmentation of chamber volumes from NCCT. Bruns et al[13] developed a DL based segmentation algorithm on NCCT for cardiac sub-structures, including chambers and myocardium, using dual energy information from dual layer detector CT scanner. This required reconstructed contrast enhanced CT scans for ground truth generation, and the corresponding virtual NCCT scans, mimicking real NCCT scans. Interestingly, this method also shows the trend of underestimation of volumes on NCCT, similar to our study. Shahzad et al[14]. developed an automatic, atlas based, cardiac segmentation algorithm to measure the volumes of the 4 cardiac chambers, the whole heart and the aortic root from non-contrast CT scans done for calcium score assessment. They showed an excellent correlation with the volumes obtained from the contrast CT. A recent study evaluated a previous version of our algorithm for automated measurement of LA volume on 273 lung cancer screening scans.[7] It showed an excellent agreement with manual quantification, with LAV intraclass correlation of 0.950 (0.936-0.960). Bland-Altman plot demonstrated the AI underestimated LAVi by a mean of 5.86 mL/m2.

Overall , these studies differed from our study in that some used either non-gated scans with differing purposes[7,15,16], geometric assumptions to infer volumes from manually traced 2D slices[4,17], performed single chamber analyses[4,17,18], required information beyond routine NCCT[13], or were done in proof-of-concept settings with small sample sizes[13,14] .

*Outlier Analyses*

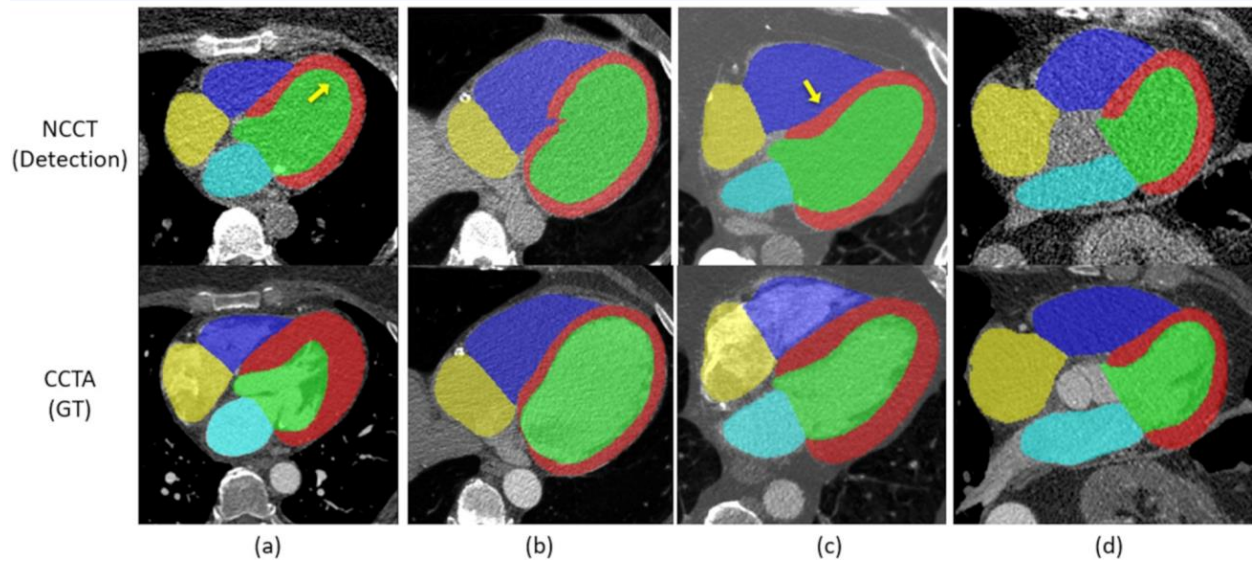

Figure 4 Modes of failure: (a) Inaccurate placement of endocardial boundary in cases with thick LV wall; (b) Cases with highly enlarged chambers; and (c) RV is overestimated due to incorrect septal wall placement and myocardium is underestimated similar to the to the case (a). d) Case with large volume difference but reasonable predicted contours.

To better understand the limitations of the method, we identified and analyzed the cases showing the highest difference in volumes. These were defined from the Bland Altman analyses as those patients where the difference between volumes from CCTA and NCCT were more than 3 standard deviations from the mean error. Among 20 patients identified, a single chamber contributed to the outlier status for 9 patients, two chambers contributed to the same for 7, and three chambers were outliers for 2 patients. For the 2 remaining patients, while none of the individual chambers were outliers, LV total volume (calculated as the sum of myocardial and LV BP volume) was the outlier. The scans and segmentation outputs of the individual patients were further analyzed to understand the reasons for the errors. Four patients had very large chamber sizes (volumes reported in Supplementary Table 2) which were captured inaccurately by the algorithm. On further investigation, these cases were found to have volumes larger than the 95$^{th}$ percentile in the training dataset used for algorithm. Augmenting the algorithm development dataset with hearts in this larger range of sizes could solve the issue.

In 10 patients, the AI algorithm was unable to place the endocardial boundary correctly because of increased LV wall thickness. Ganau et al describes two types of LV hypertrophy – eccentric hypertrophy (LV mass increases symmetrically with cavity volume), and concentric hypertrophy (increase in wall thickness outpaces the increase in cavity volume).[19] Due to the absence of any contrast in the heart, the AI algorithm learns to fit an average shape based on external image cues such as inflection points from RV apex, valve insertion points etc. While this results in reasonable predictions in cases with eccentric hypertrophy, it appears to create sub-optimal predictions in patients with concentric left ventricular hypertrophy.

One patient had the RV overestimated due to incorrect detection of the septal wall. Another patient had only partial view of the left ventricle in the NCCT scan. One patient was observed to have arrhythmia resulting in very poor image quality. For three patients, the predicted cardiac chambers on NCCT were found to be reasonable upon visual inspection, and the difference between the volumes on CCTA and NCCT are hypothesized to be due to beat-to-beat variations of cardiac volumes, or inaccuracies in the recorded cardiac phase information. Example images for the dominant failure modes are shown in Figure 4.

*Study Limitations*

Our study had some limitations. The study lacked clinical information on heart failure as well as no MRI scans were available for comparison. Although MRI is considered the gold standard for measuring chamber volumes, volumes from gated, contrast-enhanced cardiac CT and MRI have been shown to have excellent correlation.[20-21] As observed from the outlier analyses, while LV BP volumes and LV mass show high correlations on an average, the algorithm inaccurately places the endocardial border in patients with concentric LV hypertrophy. We also observe higher errors in patients with highly enlarged chambers, which can be detected in a visual review of segmentations. On a more general level, any inaccuracies in cardiac phase calculation or heart rate variability within the patient, while being hard to detect, would affect the volume comparison.

The concepts and information presented in this article are based on research results that are not commercially available. Future availability cannot be guaranteed.

## 5. Supplementary Material

*S.1 Deep Learning Model*

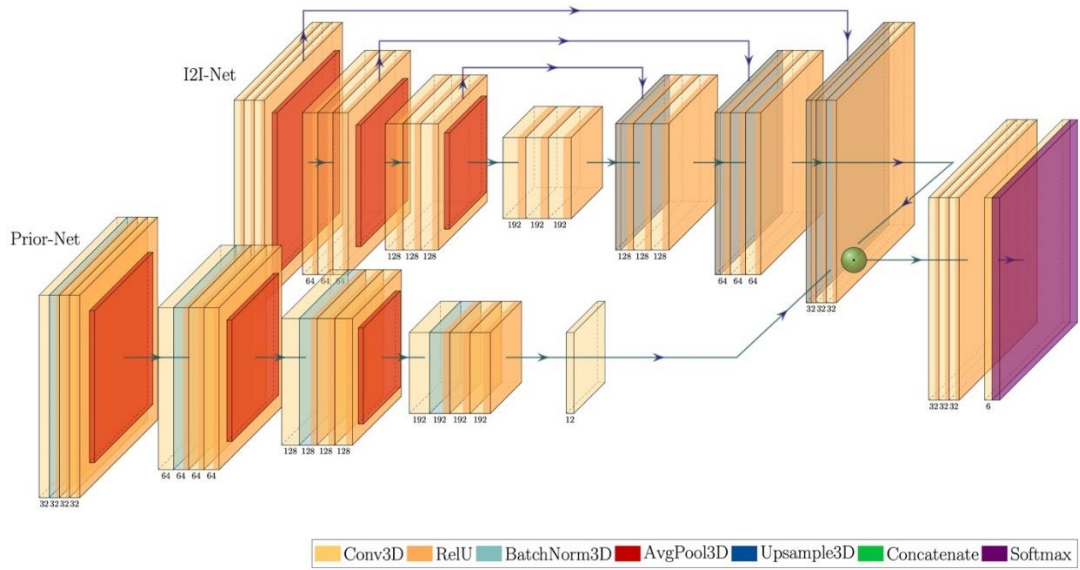

Supplementary figure 1 Algorithm architecture

The AI algorithm consisted of a 3D Image to Image (I2I) segmentation network (Supplementary figure 1), combined with a conditional Variational Autoencoder (cVAE) [8]. The algorithm was trained using multi-center, paired contrast and non-contrast scans from 2594 patients referred to CTA to diagnose coronary artery disease. Mean age in the development dataset was $60.2 \pm 11.4$ years, and 45% were women. Clinical reports were available for about half the dataset, and indicated a diverse range of structural findings, including myocardial thinning (3%), hypertrophy (11%), enlarged atria (7%), enlarged ventricles(11%), pericardial effusion (3%), and pericardial thickening (4%). The datasets were acquired using SOMATOM CT scanners (Force, Definition Flash, Definition AS+) scanners (Siemens Healthineers, Erlangen, Germany), with a variety of tube voltages ranging from 70 to 150 kVp. A patient volume may contain 41-607 slices, with slice thickness varying from 0.5 to 3 mm. The resolution within a slice is isotropic and varies from 0.25 to 0.67 mm for different volumes. The best configuration of weights and hyperparameters were chosen using a parameter tuning set of 95 patients. A separately held out testing set of 105 patients was also used to benchmark the performance of the chosen model. The training, parameter-tuning, and testing datasets were chosen randomly. All training and testing are done on 3D volumes. It is to be noted that algorithm development was performed and finalized independently, prior to the clinical validation reported in this work.

*S.2 Ground truth generation for training*

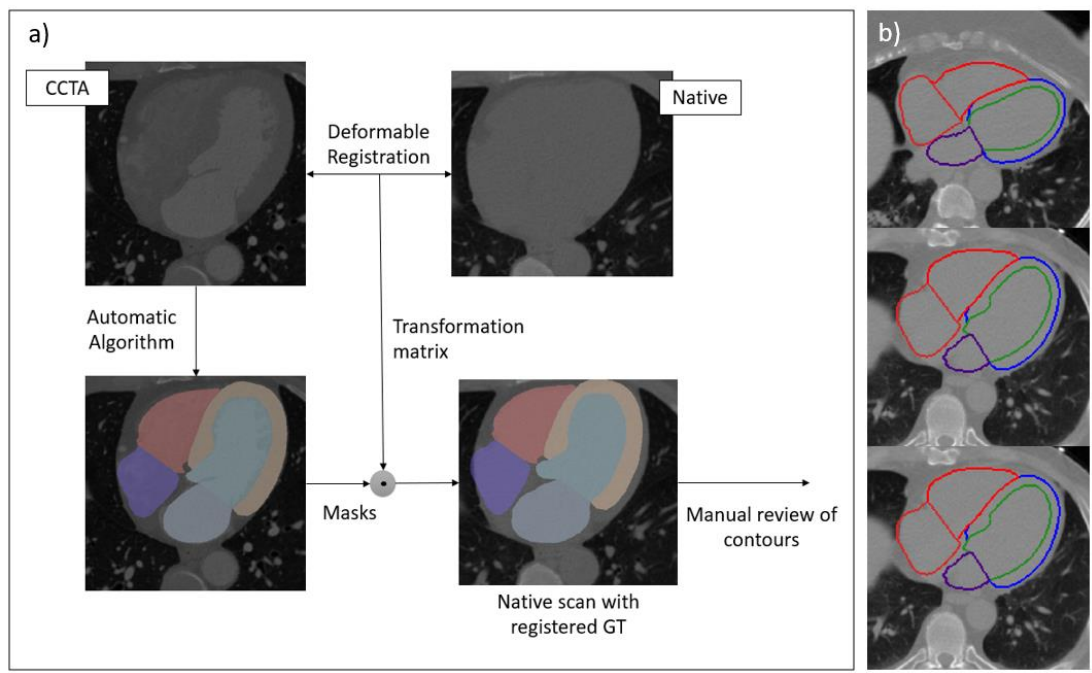

Supplementary figure 2  Pipeline for ground truth generation for algorithm development. a) Contours from closest-cardiac phase CCTA were aligned onto NCCT.  b) Example NCCTs with corresponding aligned GT contours

Contouring non-contrast CT scans often requires the expertise of cardiovascular radiologists as it is a tedious and time-consuming task, with high inter user variability. This makes it infeasible to create large-scale annotated databases of non-contrast scans to train deep learning models. We solve this problem by generating ground truth segmentations from paired contrast studies. For each non-contrast scan, we chose the corresponding CCTA scan with contrast on the same patient, that is within 20% of the cardiac phase of the non-contrast scan. The segmentation masks for four chambers and myocardium on these CCTA scans were obtained using an automated approach[9,10]. The contrasted scans were then registered using deformable registration[11] to the NCCT data. The transformation matrix was applied to the segmented masks to obtain the ground truth masks on the NCCT data for the patient. We apply this pipeline to obtain ground truth segmentations for every patient. Additionally, the obtained masks were reviewed manually to ensure high quality of the ground truth. This process of automatic generation of ground truth followed by quality review was applied to the entire database of 2794 patients used for algorithm development (Supplementary figure 2). It should be noted that this process of ground truth generation was used only for the algorithm development phase and is separate from the validation done in the current clinical study.

*S.3 Training and inference*

The input volumes for each patient underwent a simple preprocessing pipeline consisting of resampling to an isotropic 1 mm resolution, intensity truncation at 976 HU and min-max normalization to scale the input to 0-1 intensity range. Random data augmentations like translation, rotation, gaussian noise addition etc. were enabled during training to add robustness to geometric and intensity variations.

The I2I network and cVAE networks were trained simultaneously, with a combined cross entropy loss and Kullback-Leibler divergence loss[8], and L2 regularization on network weights. Such a training approach enabled the network to not only learn mapping from image to segmentation, but also to learn a latent space, encoding the joint distributions of images and possible segmentations. The network was trained from randomly initialized weights, using Kaiming normal initialization for weight matrices and truncated normal initialization for biases. Adam optimizer was used with a fixed learning rate of 1e-4. During inference, the algorithm uses only the encoder block of the cVAE network, also referred to as the prior network, and samples multiple instances to produce several plausible segmentation candidates for the input data. The final segmentation was obtained by taking the mean of these predictions. The best network weights were chosen based on highest dice score on the tuning set of 95 patient volumes. The dice similarity score measures pixel wise similarity between predicted and ground truth masks, and ranges from 0 (no overlap) to 1 (exact overlap). It should be noted that the performance metrics during training and tuning are calculated against the registered 'ground truth' from contrasted scans, which might still have minor misalignments due to the inherent difficulty of the task. Regardless, the dice score metric and volume correlations for the 105 patients in the testing set are reported in Supplementary table 1.

Supplementary table 1. Results on internal testing set. Dice metric (Mean ± SD) and volume correlations for 105 testing patients during algorithm development. Both metrics range from 0-1, and higher the better.

| Metric | LV Total | LV BP | RV | RA | LA |
|--------|----------|-------|-----|-----|-----|
| Dice | 0.86 ± 0.07 | 0.90 ± 0.06 | 0.90 ± 0.04 | 0.88 ± 0.07 | 0.93 ± 0.03 |
| Volume correlation | 0.97 | 0.85 | 0.93 | 0.91 | 0.94 |

Supplementary table 2: Predicted and GT chamber volumes for patients in outlier set with enlarged chambers. Please refer to Outlier Analyses section for more information

| Pt No. | Chamber | NCCT (Detected) (mL) | CCTA (GT) (mL) |
|---|---|---|---|
| 1 | LV Total | 639 | 757 |
|   | LV BP | 367 | 469 |
|   | LA | 150 | 186 |
| 2 | RA | 156 | 225 |
| 3 | LA | 174 | 212 |
| 4 | RV | 430 | 361 |

**References**

1. Lo-Kioeng-Shioe MS, Vavere AL, Arbab-Zadeh A, et al. Coronary Calcium Characteristics as Predictors of Major Adverse Cardiac Events in Symptomatic Patients: Insights From the CORE320 Multinational Study. *J Am Heart Assoc*. 2019;8(6). doi:10.1161/JAHA.117.007201

2. Hunt SA, Abraham WT, Chin MH, et al. ACC/AHA 2005 Guideline Update for the Diagnosis and Management of Chronic Heart Failure in the Adult: A Report of the American College of Cardiology/American Heart Association Task Force on Practice Guidelines (Writing Committee to Update the 2001 Guidelines for the Evaluation and Management of Heart Failure): Developed in Collaboration With the American College of Chest Physicians and the International Society for Heart and Lung Transplantation: Endorsed by the Heart Rhythm Society. *Circulation*. 2005;112(12). doi:10.1161/CIRCULATIONAHA.105.167586

3. Ajello L, Coppola G, Corrado E, La Franca E, Rotolo A, Assennato P. Diagnosis and Treatment of Asymptomatic Left Ventricular Systolic Dysfunction after Myocardial Infarction. *ISRN Cardiol*. 2013;2013:1-7. doi:10.1155/2013/731285

4. Bittencourt MS, Blankstein R, Mao S, et al. Left ventricular area on non-contrast cardiac computed tomography as a predictor of incident heart failure – The Multi-Ethnic Study of Atherosclerosis. *J Cardiovasc Comput Tomogr*. 2016;10(6):500-506. doi:10.1016/j.jcct.2016.07.009

17. Qureshi WT, Nasir K, Hacioglu Y, et al. Determination and distribution of left ventricular size as measured by noncontrast CT in the Multi-Ethnic Study of Atherosclerosis. *J Cardiovasc Comput Tomogr*. 2015;9(2):113-119. doi:10.1016/j.jcct.2015.01.001

18. Daniel KR, Bertoni AG, Ding J, et al. Comparison of Methods to Measure Heart Size Using Noncontrast-Enhanced Computed Tomography: Correlation With Left Ventricular Mass. *J Comput Assist Tomogr*. 2008;32(6):934-941. doi:10.1097/RCT.0b013e318159a49e

19. Ganau A, Devereux RB, Roman MJ, et al. Patterns of left ventricular hypertrophy and geometric remodeling in essential hypertension. *J Am Coll Cardiol*. 1992;19(7):1550-1558. doi:10.1016/0735-1097(92)90617-V

20. Galea N, Carbone I, Cannata D, et al. Right ventricular cardiovascular magnetic resonance imaging: normal anatomy and spectrum of pathological findings. *Insights Imaging*. 2013;4(2):213-223. doi:10.1007/s13244-013-0222-3

21. Kim JY, Suh YJ, Han K, Kim YJ, Choi BW. Cardiac CT for Measurement of Right Ventricular Volume and Function in Comparison with Cardiac MRI: A Meta-Analysis. *Korean J Radiol*. 2020;21(4):450. doi:10.3348/kjr.2019.0499
20